\documentclass[aps,prd,preprint,nobibnotes,nofootinbib]{revtex4-1}
\usepackage{amsfonts}
\usepackage{graphicx}
\usepackage{amsmath}
\usepackage{amsthm}

\newtheorem*{thm}{Theorem}

\begin{document}

\title{Motion of Small Bodies in Classical Field Theory}

\author{Samuel E. Gralla}
\affiliation{\it Enrico
Fermi Institute and Department of Physics \\ \it University of Chicago
\\ \it 5640 S.~Ellis Avenue, Chicago, IL~60637, USA}

\begin{abstract}
I show how prior work with R. Wald on geodesic motion in general relativity can be generalized to classical field theories of a metric and other tensor fields on four-dimensional spacetime that 1) are second-order and 2) follow from a diffeomorphism-covariant Lagrangian.  The approach is to consider a one-parameter-family of solutions to the field equations satisfying certain assumptions designed to reflect the existence of a body whose size, mass, and various charges are simultaneously scaled to zero.  (That such solutions exist places a further restriction on the class of theories to which our results apply.)  Assumptions are made only on the spacetime region outside of the body, so that the results apply independent of the body's composition (and, e.g., black holes are allowed).  The worldline ``left behind'' by the shrinking, disappearing body is interpreted as its lowest-order motion.  An equation for this worldline follows from the ``Bianchi identity'' for the theory, without use of any properties of the field equations beyond their being second-order.  The form of the force law for a theory therefore depends only on the ranks of its various tensor fields; the detailed properties of the field equations are relevant only for determining the charges for a particular body (which are the ``monopoles'' of its exterior fields in a suitable limiting sense).  I explicitly derive the force law (and mass-evolution law) in the case of scalar and vector fields, and give the recipe in the higher-rank case.  Note that the vector force law is quite complicated, simplifying to the Lorentz force law only in the presence of the Maxwell gauge symmetry.  Example applications of the results are the motion of ``chameleon'' bodies beyond the Newtonian limit, and the motion of bodies in (classical) non-Abelian gauge theory.  I also make some comments on the role that scaling plays in the appearance of universality in the motion of bodies.
\end{abstract}

\maketitle

\section{Introduction}

In special relativity, a non-interacting body moves in a straight line.  Therefore, it is not surprising that in general relativity an ``infinitesimal test body'' (i.e., a body small enough that the curvature of the external universe can be neglected, and weakly-gravitating enough that curvature it generates can be neglected) will move \textit{locally} in a straight line, i.e., it will follow a geodesic.  But from this perspective it does seem quite surprising that strong-field bodies like neutron stars and black holes in fact \textit{also} move on geodesics (in the limit of small size).  After all, no matter how small or light such a body, the local spacetime metric will differ significantly from that of flat spacetime, and one would therefore expect that nonlinear gravitational dynamics---certainly not special relativity---would principally determine its motion.  Furthermore, since the metrics of different strong-field bodies will differ greatly from \textit{each other}, one would perhaps expect there to be \textit{no} universal law for the motion of strong-field bodies at all.  Indeed, the natural assumption would seem to be that the motion of a strong-field body depends in detail upon its composition.

This expectation is incorrect for a very counter-intuitive reason: in general relativity, the motion of a small body is in fact completely determined by field dynamics \textit{outside} of the body.  This surprising fact was first demonstrated by Einstein, Infeld and Hoffman \cite{einstein-infeld-hoffman}, and has become the foundation of a more modern approach to motion termed ``matched asymptotic expansions'' \cite{matched-expansions} (see also \cite{poisson,gralla-wald,futamase-hogan-itoh,pound}).  The basic physical requirement of this line of work is the existence of a region (the ``buffer zone'') sufficiently far from the body that the body field may be approximated as a multipole series, yet sufficiently close to the body that the field of the external universe may be approximated in an ordinary Taylor series.  The vacuum gravitational dynamics taking place in this region then suffice to determine the motion.

A primary purpose of this paper is to determine to what extent this conclusion generalizes to other classical field theories.  To investigate this question I generalize the approach taken in \cite{gralla-wald} to deriving geodesic motion in general relativity.\footnote{I do not treat self-force corrections, which were the primary focus of \cite{gralla-wald}.}  In the formalism of \cite{gralla-wald} a small body is characterized by a one-parameter-family of solutions to the vacuum Einstein equation describing the region outside of a body that shrinks to zero size and mass with the perturbation parameter, $\lambda$.  A family with such behavior is considered by demanding the existence of a second, ``scaled'' limit wherein the coordinates and metric are rescaled such the body is held at fixed size and mass.  At $\lambda=0$ in the original limit the body disappears, leaving behind a smooth spacetime with a preferred worldline, $\gamma$, picked out; this worldline is interpreted as the lowest-order perturbative motion of the body.  We showed that $\gamma$ must be a geodesic by applying the Bianchi identity to an effective point particle description that (remarkably) emerges at first order in $\lambda$.  In this paper I generalize the approach to theories that 1) follow from a diffeomorphism-covariant Lagrangian, ensuring a ``Bianchi identity'' and 2) have second-order field equations.

  For the above class of theories the method of \cite{gralla-wald} gives an equation for $\gamma$ that depends only on ``buffer zone'' field properties, showing that the Einstein-Infeld-Hoffman idea remains correct in a more general context.  More specifically, the equation involves, in addition to the value and first-derivative of the external fields at the location of the body, various charges (understood to include mass as the charge associated with the metric) that are determined from the body's fields in the scaled limit (they are ``field monopoles'').  The results rely only on properties 1) and 2) above and are therefore surprisingly independent of the details of the theory.  In particular, the force law depends only on the form of the Bianchi identity, which in turn depends only on the ranks of the tensor fields considered (although extra identities following from gauge symmetry can greatly simplify the results).  Therefore, the expression for the force in terms of the charges and external field values is in fact identical across theories with the same types of tensor fields and the same gauge symmetries.  However, the charges associated with a particular body composition will differ in different theories, since the relationship between a given source and the field monopoles it generates will depend on the field equations.  In this way different theories will make different predictions for the motion of the ``same body,'' even when the force law is identical.

In interpreting the results it is useful to distinguish varying degrees of ``universality'' in the motion of small bodies.  In the case of general relativity, all small bodies move on geodesics, so that their internal structure is completely irrelevant to their motion.  In Einstein-Maxwell theory, a single number characterizing the body (the charge-to-mass ratio) determines how it will move, so that the internal structure is minimally relevant.  In scalar-tensor theory, a free function of time (the charge-to-mass ratio of the non-conserved charge) specifies the motion of a body, so that the internal structure is somewhat relevant.  In higher-rank theories a finite number of free functions of time characterize the motion of a body.  Of these results only geodesic motion in general relativity is truly universal in that it applies to \textit{all} bodies; however, I will refer to all of the above results as ``universal behavior in motion'', since the information required to determine the motion of a small body is reduced from the complete description of the body to the knowledge of a finite number of parameters at each time.  To adopt the language of condensed matter physics, there are thus large ``universality classes'' of small bodies that move in the same way. 
 
The content of this paper is as follows.  In section \ref{sec:GR} I summarize the formalism of \cite{gralla-wald} to derive geodesic motion in general relativity.  In section \ref{sec:scalar} I generalize the formalism to Einstein-scalar and then more general scalar-tensor theories, deriving the scalar force law.  Note that mass evolution always occurs, and the scalar charge evolution is unconstrained.  I discuss the results in the context of specific scalar-tensor theories and comment on scaling and universality.  In section \ref{sec:vector} I apply the formalism to vector-tensor theories to derive the vector force law.  This surprisingly complicated equation simplifies to the Lorentz force law in theories with the Maxwell gauge symmetry.  I also derive the simplified force law in the case of non-Abelian gauge theory.  Finally in section \ref{sec:general} I give the proof that universality in motion is achieved via buffer zone dynamics for tensor fields of arbitrary rank.  A definition and disambiguation of scale-invariance is given in an appendix.

I use the conventions of Wald \cite{wald} and work in units where $G=c=1$.  Early-alphabet Latin indices $a,b,...$ are abstract spacetime indices, while Greek indices $\mu,\nu,...$ give tensor components in a coordinate system.  When working in coordinates $(t,x^1,x^2,x^3)$, mid-alphabet Latin indices $i,j,...$ denote spatial components $1-3$, while a zero denotes the time component $t$.  Mid-alphabet capital Latin indices $I,J,...$ label members of a collection of tensor fields.

\section{Review of formalism: motion in general relativity}\label{sec:GR}

In this section I review the derivation of geodesic motion given in \cite{gralla-wald}.  While the treatment here is self-contained, the reader is referred to \cite{gralla-wald} for more details and significantly more motivation.  Note that many of the arguments given here will hold identically or analogously for the more general theories treated in later sections, in which case those arguments will not be repeated.

The basic approach to motion is to formalize the notion of a ``small body'' by considering a one-parameter-family of solutions to Einstein's equation that contains a body that shrinks to zero size with the parameter $\lambda$.  While no universal behavior in motion (nor even any definition of ``position''\footnote{Consider, for example, the impossible task of assigning a center of mass position to a black hole.}) can possibly be obtained at any finite $\lambda$, in the limit $\lambda \rightarrow 0$ one can hope for a simplified description, whose observables will then approximate observables at small but finite $\lambda$.\footnote{Of course, the value of $\lambda$ at which the physical spacetime is embedded into the one-parameter-family is arbitrary.  What matters for the application of the simplified description is that corrections to the relevant observables are numerically small.}  The task is therefore to develop assumptions on a one-parameter-family to the effect that it contains a body shrinking to zero size.  The first realization is that the body must also shrink to zero mass, since (roughly speaking) no body can be smaller than its Schwarzschild radius.  The body will thus disappear in the limit, but it will leave behind a preferred worldline, $\gamma$, characterizing its motion.  Our method of considering such a body is essentially to demand that if we zoom in on the presumed shrinking body, then a body is recovered.  This zooming process is accomplished via the notion of a \textit{scaled limit}, defined as follows.  Consider a one-parameter-family of metrics $g_{ab}(\lambda)$, whose metric components $g_{\mu \nu}(\lambda)$ are given in some particular coordinates $(t,x^i)$.  Introduce both a rescaled metric $\bar{g}_{ab}(\lambda) = \lambda^{-2} g_{ab}(\lambda)$ and, for a particular time $t_0$, rescaled coordinates $(\bar{t},\bar{x}^i)=\left((t-t_0)/\lambda,x^i/\lambda\right)$.  Then, the scaled limit is given by 
\begin{equation}\label{eq:scaled-limit}
\bar{g}^{(0)}_{\bar{\mu} \bar{\nu}}(t_0) \equiv \lim_{\lambda \rightarrow 0}\bar{g}_{\bar{\mu} \bar{\nu}}(\lambda;t_0),
\end{equation}
where the limit is taken at fixed scaled coordinate.  In this notation, the bar on the ``$g$'' indicates that the rescaled metric is being considered, while the bars on the coordinate component indices ``$\mu$'' and ``$\nu$'' indicate that the components of $\bar{g}_{ab}$ in the rescaled coordinates are being considered.  (I will continue to adopt this notation throughout the paper.\footnote{Note also that coordinate indices will \textit{always} refer to the original cartesian-like $(t,x^i)$, even if the coordinate components are being viewed as functions of other variables, such as spherical coordinates.})

  This limit has the interpretation of ``zooming in'' because a fixed-$\bar{x}^\mu$ observer moves ever closer to the shrinking body while the rescaled metric keeps distances finite.  A simple example to keep in mind is the family of Schwarzschild deSitter metrics of mass parameter $\lambda$ \cite{gralla-wald}, which clearly contains a shrinking body of the sort we want to consider.  In the ordinary limit the body disappears, leaving behind the background spacetime of deSitter.  But in the scaled limit the background ``disappears'', leaving behind the Schwarzschild metric for all $t_0$.  For all one-parameter-families in this paper, I will refer to $g^{(0)}_{\mu \nu} \equiv g_{\mu \nu}(\lambda=0)$ as the ``background metric'' and to $\bar{g}^{(0)}_{\bar{\mu} \bar{\nu}}(t_0)$ as the ``body exterior metric'' (no interior is ever considered).  The existence of both original and scaled limits reflects the presence of a body with the appropriate scaling, and will be our assumptions i) and ii), below.

An additional assumption is required.  To arrive at this assumption, note that the rescaling of the metric by $\lambda^{-2}$ effectively cancels powers of $\lambda^2$ that arise in changing to the rescaled coordinates, so that one has the simple formula
\begin{equation}\label{eq:compute-scaled}
\bar{g}_{\bar{\mu} \bar{\nu}}(\lambda;t_0;\bar{t},\bar{x}^i) = g_{\mu \nu}(\lambda;t=t_0+\lambda \bar{t}, x^i=\lambda \bar{x}^i),
\end{equation}
relating barred components of the barred metric, $\bar{g}_{\bar{\mu} \bar{\nu}}$, to corresponding unbarred coordinates of the unbarred metric, $g_{\mu \nu}$.  That is, one simply ``plugs in'' $t=t_0+\lambda \bar{t}$ and $x^i=\lambda \bar{x}^i$ to compute $\bar{g}_{\bar{\mu} \bar{\nu}}$ from $g_{\mu \nu}$.  This formula shows that if we regard $g_{\mu \nu}$ as a function of new variables $\alpha\equiv r$,$\beta\equiv \lambda/r$,$\theta$,$\phi$ (at fixed $t$, and with $r,\theta,\phi$ defined relative to $x^i$ in the usual way), then the scaled limit \eqref{eq:scaled-limit} is given by the limit $\alpha \rightarrow 0$ at fixed $\beta$ of the original metric components $g_{\mu \nu}$.  Similarly, the original limit is given by the limit $\beta \rightarrow 0$ at fixed $\alpha$.  Demanding that both limits exist is thus the statement of separate continuity in $\alpha$ and $\beta$.  A natural extension (argued for at length in \cite{gralla-wald} on the grounds that it excludes certain pathological behavior) is to demand \textit{joint} continuity and in fact joint smoothness in $\alpha$ and $\beta$ (although only $C^1$ is required here).  This will be our assumption iii), below.  Note that the electromagnetic analog of iii) has been shown to hold for a family of shrinking charge-current sources in flat spacetime \cite{gralla-harte-wald}.

The assumptions of this section for the metric family $g_{ab}(\lambda)$ are that there exists coordinates $(t,x^i)$ at each $\lambda \geq 0$ such that the following hold.
\begin{itemize}
\item i) For $r \geq \lambda \bar{R}$ for some constant $\bar{R}$, the metric components $g_{\mu \nu}(\lambda)$ satisfy the vacuum Einstein equation and are smooth functions of $(t,x^i,\lambda)$.  The worldline, $\gamma$, defined by $\lambda = x^i = 0$ is timelike.
\item ii) The scaled metric components $\bar{g}_{\bar{\mu} \bar{\nu}}(t_0;\lambda)$ are smooth functions of $(\lambda,\bar{t},\bar{x}^i)$ for $\bar{r} \geq \bar{R}$.
\item iii)  The metric components $g_{\mu \nu}$ are smooth functions of $(\alpha,\beta)$ at $(0,0)$ for fixed $(t,\theta,\phi)$.
\end{itemize}
Assumption i) establishes our domain $r \geq \lambda \bar{R}$ and provides the requisite smoothness for perturbation theory on that domain.  It also lays the groundwork for the interpretation of the domain as the exterior of a shrinking body by taking $\gamma$ to be timelike.  Assumption ii) establishes this interpretation according to the ideas of the scaled limit, and assumption iii) adds additional ``uniformity'' properties \cite{gralla-wald}.  Note, however, that this latter assumption has an important physical consequence (i.e., it places an important restriction on the type of spacetime for which our approximate results will be useful).  It requires that there be a spatial region both far enough from the body that its field can be approximated in a series in inverse powers of distance (corrections in $\beta$ near zero), and close enough to the body that the field of the external universe can be approximated as a series in positive powers of distance (corrections in $\alpha$ near zero).  Therefore by seeking one-parameter-families containing a shrinking body, we in fact end up with a mathematically precise version of the usual ``buffer zone'' assumption of the Einstein-Infeld-Hoffman approach and its descendants.  No further assumptions beyond i), ii), iii) (and their analogs for other theories) are made in this paper.

Smoothness in $\alpha$ and $\beta$ allows us to Taylor expand in these variables to any finite order.  However, to derive geodesic motion we in fact require only a single derivative in $\beta$,
\begin{align}
g_{\mu \nu}(\lambda,t,r,\theta,\phi) & = b_{\mu \nu}(t,\theta,\phi) + c_{\mu \nu}(t,\theta,\phi) \beta + O(\alpha) + O(\beta^2) \nonumber \\
& = b_{\mu \nu}(t,\theta,\phi) + c_{\mu \nu}(t,\theta,\phi) \frac{\lambda}{r} + O(r) + O\left(\frac{\lambda}{r}\right)^2  \label{eq:galphabeta}
\end{align}
where $O(r)$ near zero is at fixed $\lambda/r$, and $O(\lambda/r)$ near zero is at fixed $r$.  Sorting into powers of $\lambda$ and $r$, we have
\begin{equation}
g_{\mu \nu}(\lambda,t,r,\theta,\phi) = b_{\mu \nu}(t,\theta,\phi) + O(r) + \lambda \left( c_{\mu \nu}(t,\theta,\phi)\frac{1}{r} + O(r^0) \right) + O(\lambda^2), \qquad r>0
\end{equation}
where the order symbols are for small $r$ and $\lambda$.  From this expression it is easy to read off series expressions for the background metric $g^{(0)}_{\mu \nu}$ and linear perturbations $g^{(1)}_{\mu \nu} \equiv \partial_\lambda g_{\mu \nu}|_{\lambda=0}$,
\begin{align}
g^{(0)}_{\mu \nu}(t,x^i) & = b_{\mu \nu}(t) + O(r), & r\geq 0 \ \label{eq:g0} \\
g^{(1)}_{\mu \nu}(t,x^i) & = c_{\mu \nu}(t,\theta,\phi)\frac{1}{r} + O(1), & r>0. \label{eq:g1} 
\end{align}
Since $g^{(0)}_{\mu \nu}$ is assumed smooth everywhere, $b_{\mu \nu}$ cannot depend on angles and we have written $b_{\mu \nu}(t)$.  (This quantity is usually taken to be $\eta_{\mu \nu}$ by coordinate choice.)  We are also interested in the consequences of equation \eqref{eq:galphabeta} in the scaled limit.  Using equation \eqref{eq:compute-scaled}, we have
\begin{equation}
\bar{g}_{\bar{\mu} \bar{\nu}}(\lambda;t_0;\bar{t},\bar{r},\theta,\phi) = b_{\mu \nu}(t_0 + \lambda \bar{t},\theta,\phi) + c_{\mu \nu}(t_0 + \lambda \bar{t},\theta,\phi) \frac{1}{\bar{r}} + O(\lambda \bar{r}) + O\left(\frac{1}{\bar{r}^2}\right),
\end{equation}
so that the limit $\lambda \rightarrow 0$ at fixed $(\bar{t},\bar{x}^i)$ gives
\begin{align}
\bar{g}^{(0)}_{\bar{\mu} \bar{\nu}}(t_0;\bar{x}^i) & = b_{\mu \nu}(t_0) + c_{\mu \nu}(t_0,\theta,\phi) \frac{1}{\bar{r}} + O\left(\frac{1}{\bar{r}^2}\right),  \label{eq:gbar0}
\end{align}
and the body exterior metric $\bar{g}^{(0)}_{\bar{\mu} \bar{\nu}}$ is seen to be \textit{stationary} (independent of $\bar{t}$) and \textit{asymptotically flat} (constant as $\bar{r} \rightarrow \infty$).  (Stationarity follows from smoothness of $g_{\mu \nu}(\lambda)$ in $t$.)  This supports the idea that $\bar{g}^{(0)}_{\bar{\mu} \bar{\nu}}$ characterizes the body exterior as it would appear in isolation at time $t_0$.  Equations \eqref{eq:g0}, \eqref{eq:g1}, and \eqref{eq:gbar0} are key consequences of our assumptions.

Geodesic motion is now derived as follows.  By assumption i), $g_{\mu \nu}(\lambda)$ is smooth in $\lambda$ at $\lambda=0$ for $r>0$.  Thus $g^{(1)}_{\mu \nu}$ satisfies the vacuum linearized Einstein equation about $g^{(0)}$,
\begin{equation}\label{eq:G1}
G^{(1)}_{\mu \nu}[g^{(1)}] = 0 , \qquad r > 0.
\end{equation}
Now regard $g^{(1)}_{\mu \nu}$ as a distribution on defined on the background spacetime including at $r=0$, which is possible because its ``most singular'' behavior is only $1/r$.  Since $G^{(1)}$ is a second-order, linear partial differential operator, it follows from equations \eqref{eq:g1} and \eqref{eq:G1} via the analysis of appendix \ref{sec:calc} that, distributionally, we have
\begin{equation}\label{eq:G1dist}
G^{(1)}_{\mu \nu}[g^{(1)}] = N_{\mu \nu}(t) \delta^{(3)}(x^i),
\end{equation}
for some $N_{\mu \nu}$ defined on the worldline $x^i=0$.  (This result is analogous to the well-known fact that $\nabla^2(1/r)=- 4 \pi \delta^{(3)}(x^i)$.  If the explicit form of $G^{(1)}_{\mu \nu}$ is used, a formula for may be obtained for $N_{\mu \nu}$ in terms of angle averages of $c_{\mu \nu}$ and its first angular derivatives.) 
Thus, an effective distributional stress-energy of $1/8\pi N_{\mu \nu} \delta^3(x^i)$ has emerged at first-order in perturbation theory, supported on the worldline $\gamma$.  This is remarkable, given that any true stress-energy associated with the body is confined to $r<\lambda \bar{R}$ and excluded from consideration; and furthermore, the body need not be ``made'' of stress-energy at all (as in the case of a black hole).  

The strategy is now to apply ``conservation'' to the ``stress-energy''.  That is, because the linearized Bianchi identity $\nabla^a G^{(1)}_{ab}[g^{(1)}]=0$ holds as an identity on \textit{all} sufficiently smooth $g^{(1)}_{ab}$ (not necessarily satisfying the linearized Einstein equation), the distributional linearized Bianchi identity also holds as an identity on distributional $g^{(1)}_{ab}$, and we must have
\begin{equation}\label{eq:dist-cons}
\nabla^{\mu} \left( N_{\mu \nu}(t) \delta^{(3)}(x^i) \right) = 0
\end{equation}  
in the distributional sense.  Here $\nabla_a$ is the derivative operator associated with the background metric $g^{(0)}_{ab}$.  The consequences of this equation can be determined in a variety of ways.  I will proceed by adopting the specific coordinate choice of Fermi normal coordinates (see, e.g., \cite{poisson}) for the background metric $g^{(0)}_{\mu \nu}$.  On the worldline $x^i=0$, the metric components are Minkowski ($g_{\mu \nu}=\eta_{\mu \nu}$), and the Christoffel symbols are given by $\Gamma^i_{00} = \Gamma^0_{0i}=a_i$, where $a_i$ are the spatial components of the four-acceleration of the worldline (the time component $a_0$ is zero).  We then have
\begin{align}
\nabla^{\mu} \left( N_{\mu 0} \delta^{(3)}(\vec{x}) \right) & = \delta^{(3)}(\vec{x})\left[ - \partial_0 N_{00} + a_i N_{i0} \right] + \partial_i \delta^{(3)}(\vec{x}) \left[ N_{i0} \right] \\
\nabla^{\mu} \left( N_{\mu i} \delta^{(3)}(\vec{x}) \right) & = \delta^{(3)}(\vec{x})\left[ - \partial_0 N_{0i} + a_j N_{ij} + a_i N_{00} \right] + \partial_i \delta^{(3)}(\vec{x}) \left[ N_{ij} \right],
\end{align}
with repeated spatial indices summed.  The coefficients of $\delta^{(3)}(\vec{x})$ and $\partial_i \delta^{(3)}(\vec{x})$ must separately vanish, giving $N_{i0}=N_{ij}=0$, as well as $\partial_0 N_{00}=0$ and $a_i N_{00}=0$, i.e., geodesic motion when $N_{00}$ is non-zero.  

We can interpret $N_{00}$ through its appearance in equation \eqref{eq:G1dist}.  Since it multiplies the delta function, it is clear that $N_{00}$ will determine the singular behavior of the metric perturbation, i.e., it will determine the coefficient $c_{\mu \nu}$ in equation \eqref{eq:g1}.  Furthermore, if one expands the background metric according to \eqref{eq:g0}, it is clear that only the constant term $b_{\mu \nu}$ (which here equals $\eta_{\mu \nu}$ by coordinate choice) is relevant for the determination of $c_{\mu \nu}$ via \eqref{eq:G1dist}.  Thus, we may compute this coefficient by using the stationary linearized Einstein equation off of flat spacetime in global inertial coordinates.  The solutions are well known and one obtains for the time-time component (which is all we need) that $4 \pi c_{00} = N_{00}$ (so that $c_{00}$ is in fact independent of angles in these coordinates).  Observing the appearance of $c_{\mu \nu}$ in the series for the body exterior metric \eqref{eq:gbar0}, we conclude that $N_{00}$ is $8 \pi$ times the \textit{ADM mass} of the body exterior metric.  Therefore we define $M=(1/8 \pi) N_{00}$ and refer to this quantity as the mass of the body.  This explains the role of the requirement that $N_{00} \neq 0$ for geodesic motion to hold: there must actually be a body present in the one-parameter-family for the curve to be necessarily geodesic.

Equation \eqref{eq:G1dist} for the effective stress-energy may be clarified by introducing $M$ and by rewriting in covariant form.  Since $u^\alpha=(1,\vec{0})$ in Fermi normal coordinates, we have from $N_{ij}=N_{i0}=0$ that $N_{ab} = M u_a u_b$.  The spatial coordinate delta function becomes a worldline integral of the ``invariant'' four-dimensional delta function $\delta_4(x,x')=\frac{\delta^{(4)}(x^\mu-x'^\mu)}{\sqrt{-g}}$.  Thus we have
\begin{equation}\label{eq:pp}
G^{(1)}_{ab}[g^{(1)}] = 8 \pi M \int_\gamma u_a u_b \delta_4(x,z(\tau)) d\tau ,
\end{equation}
where the mass $M$ is constant.  This equation shows that the metric perturbations for our family are in fact sourced by the usual ``point particle'' stress-energy (see, e.g., \cite{poisson}).  Thus, despite the fact that point particles do not make sense in general relativity \cite{geroch-traschen}, we have shown that they do emerge as part of a (mathematically rigorous) approximate description of the metric of an arbitrary small body.  Furthermore, the ``particle mass'' $M$ is indeed the ADM mass of the body (as measured in the scaled limit).

The results of this section (i.e., the results of sec. IV of \cite{gralla-wald}) may be summarized as follows.  Consider a one-parameter-family of spacetimes containing a body whose size and mass decrease to zero, according to the stated assumptions.  Then, the ADM mass $M$ of the body exterior metric is a constant independent of time $t_0$, and, if $M\neq 0$, the worldline $\gamma$ left behind after the body disappears is a geodesic of the spacetime $g^{(0)}_{ab}$ left behind.  Furthermore, the far-field effective description in linearized gravity is that of a point particle of mass $M$.  These results show, in essence, that small bodies move on geodesics while keeping their ADM mass constant and sourcing linear perturbations reflecting a point particle of that mass.

\section{Scalar-tensor theories}\label{sec:scalar}

A simple generalization of general relativity is the addition of a scalar field.  I will first consider the ordinary Einstein-scalar theory in detail.  I will then discuss the general case, which in fact follows from the computations already done.  Finally I make some comments on scaling and universality.  In this and later sections it will be convenient to use a Lagrangian formulation.  I will use the definitions and conventions of appendix E of Wald \cite{wald}, except that I will denote his fixed volume element $\mathbf{e}$ by ``$d^4x$''.

\subsection{Einstein-scalar theory}

The action for general relativity plus a minimally-coupled massless scalar field $\phi$ is given by
\begin{equation}\label{eq:es-action}
S = \int  d^4x \sqrt{-g} \left( R - 2 g^{ab} \nabla_a \phi \nabla_b \phi \right),
\end{equation}
where $R$ is the Ricci scalar constructed from $g_{ab}$.  I have chosen the relative normalization so that the theory reduces to that of Quinn \cite{quinn} in the appropriate limit.  It is helpful to define ${E}^{[g]}_{ab} \equiv (-g)^{-1/2} \delta S / \delta g^{ab}$ and $E^{[\phi]} \equiv (-g)^{-1/2} \delta S / \delta \phi$, which evaluate to
\begin{align}
E^{[g]}_{ab} & = G_{ab} - 2 \left( \nabla_a \phi \nabla_b \phi - \frac{1}{2} g_{ab} g^{cd} \nabla_c \phi \nabla_d \phi \right) \label{eq:Eg}\\
E^{[\phi]} & = 4 g^{ab} \nabla_a \nabla_b \phi . \label{eq:Ephi}
\end{align}
The equations of motion for the Einstein-scalar theory are then simply 
$E^{[g]}_{ab} = 0$ and $E^{[\phi]} = 0$.

The formalism requires a ``Bianchi identity'' for this theory.  To derive 
such an identity, consider the variation of the action \eqref{eq:es-action} with respect to an infinitessimal diffeomorphism.  Since the action is diffeomorphism-invariant, the variation must vanish, and one has
\begin{equation}
0 = \int d^4x \sqrt{-g} \left\{ E^{[g]}_{ab} (-2\nabla^a \xi^b) + E^{[\phi]} \xi^a \nabla_a \phi \right\},
\end{equation}
where $\xi^a$ is an arbitrary vector field.  Integrating the first term by parts, we derive
\begin{equation}\label{eq:es-bianchi}
\nabla^a E^{[g]}_{ab} = -\frac{1}{2} E^{[\phi]} \nabla_b \phi.
\end{equation}
The field equations were not used in deriving this equation, which therefore holds as an identity on all sufficiently smooth $\{ g_{ab},\phi \}$.  (This can also be easily checked by direct calculation using equations \eqref{eq:Eg} and \eqref{eq:Ephi}.)  To interpret this identity, note that nonzero values of $E^{[g]}_{ab}$ and $E^{\phi}$ would normally be interpreted as stress-energy and scalar charge density (respectively) associated with some matter field.   This equation gives the precise non-conservation the matter stress-energy in terms of the matter scalar charge density necessary for consistent coupling of that matter to the Einstein-scalar theory.  Although we will always impose $E^{[g]}_{ab}=0$ and $E^{\phi}=0$ at finite $\lambda$, non-zero values will emerge in the linearized, distributional description (analogously to equation \eqref{eq:G1dist} in general relativity), reflecting an effective stress-energy and scalar charge of the body.

We now seek to generalize the assumptions used in general relativity to the Einstein-scalar theory.  The main requirement is to take the scalar charge to zero along with the size and the mass, in order to keep the energy in the field finite.  Thus we seek a one-paramater family with scalar field behavior like $\phi \sim \lambda/r$.  To characterize this by the existence of a scaled limit, the appropriate rescaling (after changing to scaled coordinates) is simply $\bar{\phi}=\phi$.  (That is, no rescaling is required; however, we still define $\bar{\phi}$ for notational consistency.)  Then for an arbitrary family we define the scaled limit as in \eqref{eq:scaled-limit},
\begin{equation}\label{eq:scaled-limit-phi}
\bar{\phi}^{(0)} \equiv \lim_{\lambda \rightarrow 0}\bar{\phi}(\lambda), 
\end{equation}
where the limit is taken at fixed scaled coordinate.  I will refer to $\bar{\phi}^{(0)}$ as the body exterior scalar field in analogy with the body exterior metric $\bar{g}^{(0)}_{\bar{\mu} \bar{\nu}}$.

One can now follow the same path of reasoning as in section \ref{sec:GR}, leading one to assume the existence of original (i) and scaled (ii) limits, as well as the uniformity condition (iii).  In other words, the appropriate assumptions for this section are those of \ref{sec:GR}, where the metric $g_{ab}(\lambda)$ is replaced by the pair $\{g_{ab}(\lambda),\phi(\lambda)\}$, and the required equations are not Einstein's equation but the Einstein-scalar equations $E^{[g]}_{ab} = 0$ and $E^{[\phi]} = 0$.  (The same coordinates $(t,x^i)$ and hence worldline $\gamma$ are used for the metric and scalar field.)  Note that the new assumption iii) will imply the existence of a ``buffer zone'' for the scalar field as well as the metric.

The steps of the derivation of motion now follow completely analogously.  The analog of \eqref{eq:galphabeta} holds for the scalar field, which leads to the analogs of \eqref{eq:g0}, \eqref{eq:g1} and \eqref{eq:gbar0}, given by
\begin{align}
\phi^{(0)} & = b^{[\phi]}(t) + O(r) \label{eq:phi0} \\
\phi^{(1)} & = c^{[\phi]}(t,\theta,\phi)\frac{1}{r} + O(1) \label{eq:phi1}\\
\bar{\phi}^{(0)} & = b^{[\phi]}(t_0) + c^{[\phi]}(t_0,\theta,\phi) \frac{1}{\bar{r}} + O \left( \frac{1}{\bar{r}^2}\right) \label{eq:phibar0}
\end{align}
for some $b^{[\phi]}$ and $c^{[\phi]}$.  (Here $\phi^{(1)}\equiv \partial_\lambda \phi|_{\lambda=0}$ is the scalar field perturbation, so that one has $\phi(\lambda) = \phi^{(0)} + \lambda \phi^{(1)} + O(\lambda^2)$ for $r>0$.)  The body exterior scalar field  $\bar{\phi}^{(0)}$ is seen to be stationary and to have smooth behavior in $1/\bar{r}$ as $\bar{r} \rightarrow \infty$, supporting the name we have given it.  Note, however, that in contrast to the case of the metric, the body exterior scalar field retains some ``memory'' of the external universe, since $b^{[\phi]}$ contains physical information about $\phi^{(0)}$.  As before, the perturbations $\phi^{(1)}$ and $g^{(1)}_{\mu \nu}$ satisfy the linearized field equations at $r>0$,
\begin{align}
E^{[g](1)}_{\mu \nu}[g^{(1)},\phi^{(1)}] & = 0, \qquad & r>0 \\
E^{[\phi](1)}[g^{(1)},\phi^{(1)}] & = 0, \qquad & r>0
\end{align}
where $E^{[g](1)}_{ab}$ and $E^{[\phi](1)}$ denote the linearizations of $E^{[g]}_{ab}$ and $E^{[\phi]}$, respectively, off of the background $\{g^{(0)},\phi^{(0)}\}$.  Since the field equations are second order in $g_{ab}$ and in $\phi$, the linear operators $E^{[g](1)}_{ab}$ and $E^{[\phi](1)}$ are also second order in $g^{(1)}_{ab}$ and $\phi^{(1)}$.  Thus, as in the case of pure gravity, the ``$1/r$ behavior'' of equations \eqref{eq:g1} and \eqref{eq:phi1} implies via the analysis of appendix \ref{sec:calc} that, distributionally, we have
\begin{align}\label{eq:Eg1}
E^{[g](1)}_{\mu \nu}[g^{(1)},\phi^{(1)}] & = N^{[g]}_{\mu \nu}(t) \delta^{(3)}(x^i) \\
E^{[\phi](1)}[g^{(1)},\phi^{(1)}] & = N^{[\phi]}(t) \delta^{(3)}(x^i),\label{eq:Ephi1}
\end{align}
for some $N^{[g]}_{\mu \nu}$ and $N^{[\phi]}$.  As discussed above, a non-zero value of $E^{[g]}_{\mu \nu}$ would normally be interpreted as a  matter stress-energy source, whereas a non-zero value of $E^{[\phi]}$ would be interpreted as a scalar charge density source.  Thus, despite the fact that no matter stress-energy or scalar charge density was considered, distributional descriptions of both have arisen effectively in perturbation theory.  We can take advantage of this remarkable occurrence, as before, by using the ``Bianchi identity'' for the theory.  That is, since the linearization of equation \eqref{eq:es-bianchi}, $\nabla^\mu E^{[g](1)}_{\mu \nu} = - 1/2 E^{[\phi](1)} \nabla_\nu \phi^{(0)}$, holds as an identity on all sufficiently smooth $\{g^{(1)}_{\mu \nu},\phi^{(1)}\}$, it must hold as an identity on distributions as well.  Thus we must have
\begin{equation}\label{eq:es-bianchi2}
\nabla^\mu \left( N^{[g]}_{\mu \nu} \delta^{(3)}(x^i) \right) = - \frac{1}{2} N^{[\phi]} \delta^{(3)}(x^i) \nabla_\nu \phi^{(0)}
\end{equation}
in the distributional sense.  Here $\nabla_a$ is the derivative operator associated with the background metric $g^{(0)}_{ab}$.  Adopting Fermi normal coordinates as before, this becomes
\begin{align}
0 & = \delta^{(3)}(\vec{x})\left[ - \partial_0 N^{[g]}_{00} + a_i N^{[g]}_{i0} + \frac{1}{2} N^{[\phi]} \partial_0 \phi^{(0)} \right] + \partial_i \delta^{(3)}(\vec{x}) \left[ N^{[g]}_{i0} \right] \\
0 & = \delta^{(3)}(\vec{x})\left[ - \partial_0 N^{[g]}_{0i} + a_j N^{[g]}_{ij} + a_i N^{[g]}_{00} + \frac{1}{2} N^{[\phi]} \partial_i \phi^{(0)} \right] + \partial_i \delta^{(3)}(\vec{x}) \left[ N^{[g]}_{ij} \right],
\end{align}
from which we determine $N^{[g]}_{i\mu} = 0$ as well as 
\begin{align}
\partial_0 N^{[g]}_{00} & = \frac{1}{2} N^{[\phi]} \partial_0 \phi^{(0)}|_{x^i=0} \label{eq:eom-m-coord} \\ N^{[g]}_{00} a_i & = - \frac{1}{2} N^{[\phi]} \partial_i \phi^{(0)}|_{x^i=0}. \label{eq:eom-gamma-coord}
\end{align}
We can interpret $N^{[g]}_{00}(t)$ and $N^{[\phi]}(t)$ as follows.  The formula for $E^{[g]}_{ab}$, equation \eqref{eq:Eg}, demonstrates that the linearization, $E^{[g](1)}_{ab}$ will depend on second-derivatives of $g^{(1)}_{ab}$ only through the linearized Einstein tensor $G^{(1)}_{ab}$, and will contain no second-derivatives of $\phi^{(1)}$.  Therefore, the identical arguments from the case of pure gravity carry over, and we have $(1/4\pi)N^{[g]}_{00}=c_{00}=2M$, where $M$ is the mass\footnote{Applying the usual notion of mass to Einstein-scalar theory makes sense because the scalar field stress-energy is quadratic in first-derivatives of $\phi$, so that the $1/\bar{r}$ part of the metric still satisfies the same equations as it does in general relativity.} of the body exterior metric at time $t_0=t$.  Similarly, from the properties of the stationary, flat-spacetime Klein-Gordon equation in global inertial coordinates with delta-function source (i.e., the Green's function for the Poisson equation), we have $N^{[\phi]}=-16 \pi c^{[\phi]}$; and since $c^{[\phi]}$ is the coefficient of $1/\bar{r}$ in equation \eqref{eq:phibar0}, it is in fact just the \textit{scalar charge} $q$ of the body exterior scalar field $\bar{\phi}^{(0)}$; therefore we take $N^{[\phi]}=- 16 \pi q$.

Incorporating $M$ and $q$ and rewriting equations (\ref{eq:Eg1}-\ref{eq:Ephi1}) and (\ref{eq:eom-m-coord}-\ref{eq:eom-gamma-coord}) in covariant\footnote{Note that spatial components of a tensor $T_i$ in Fermi normal coordinates correspond to projections orthogonal to $\gamma$, $(\delta_a^{ \ b} + u_a u^b )T_b$.} form, we have
\begin{align}
E^{[g](1)}_{ab} & = 8 \pi \int_\gamma M u_a u_b \delta_4(x,z(\tau)) d\tau \label{eq:E1ab}\\
E^{[\phi](1)} & = - 16 \pi \int_\gamma q \delta_4(x,z(\tau)) d\tau , \label{eq:E1}
\end{align}
as well as
\begin{align}
M u^b \nabla_b u^a & = q \left( g^{ab} + u^a u^b \right) \nabla_b \phi^{(0)} \label{eq:eom-gamma} \\
u^a \nabla_a M & = - q u^a \nabla_a \phi^{(0)}, \label{eq:eom-m}
\end{align}
where $\nabla_a\phi^{(0)}$ is evaluated on $\gamma$.  Equations \eqref{eq:E1ab} and \eqref{eq:E1} give the metric and scalar perturbations produced by the body, showing that they are in fact sourced by the usual point charge stress-energy and scalar charge density \cite{quinn,poisson}.  Equations \eqref{eq:eom-gamma} and \eqref{eq:eom-m} give the worldline and mass evolution, and agree with the equations normally given for scalar charges \cite{quinn,poisson}.  Note that the charge $q$ is unconstrained; a separate postulate about the body---such as constant charge or some other evolution law for $q$---is required to obtain a deterministic set of equations.  (Since there is no ``conservation law'' for $q$, the body can modify it via internal dynamics at will.)  Thus the possible motions small bodies in Einstein-scalar theory are specified by one free function of time.  Equations (\ref{eq:E1ab}-\ref{eq:eom-m}) give the universal behavior of small bodies in Einstein-scalar theory and comprise the results of this subsection.

\subsection{More General Scalar-Tensor Theories}

The analysis of the preceding subsection carries over straightforwardly to many more general scalar-tensor theories.  In fact, the analysis \textit{already applies} to the majority of scalar-tensor theories commonly considered, since these theories have an ``Einstein frame'' (i.e., a field redefinition) in which the matterless Lagrangian (all we ever consider) reduces precisely to equation \eqref{eq:es-action}.  However, suppose that a ``Jordan frame'' derivation is desired, or that one considers a scalar-tensor theory with no Einstein frame.  In fact, the analysis still carries over to these cases with essentially no modification.  Specifically, consider an action 
\begin{equation}\label{eq:scalar-action}
S = \int d^4x \mathcal{L}(g^{ab},\phi)
\end{equation}
such that $\mathcal{L}(g^{ab},\phi)$ is diffeomorphism-covariant\footnote{i.e., such that $\mathcal{L}$ satisfies $\mathcal{L}(\psi_*g^{ab},\psi_*\phi) = \psi_*\mathcal{L}(g^{ab},\phi)$ for diffemorphisms $\psi$} and such that ${E}^{[g]}_{ab} \equiv (-g)^{-1/2} \delta S / \delta g^{ab}$ and $E^{[\phi]} \equiv (-g)^{-1/2} \delta S / \delta \phi$ are second-order (local) differential operators on $\{ g_{ab}, \phi \}$.  The Bianchi identity is again given by equation \eqref{eq:es-bianchi}, where now ${E}^{[g]}_{ab}$ and $E^{[\phi]}$ refer to the new action \eqref{eq:scalar-action}.  Now adopt the same assumptions for the one-parameter-family $\{ g_{ab}(\lambda), \phi(\lambda) \}$, leading to equations (\ref{eq:phi0}-\ref{eq:phibar0}).  Since ${E}^{[g]}_{ab}$ and $E^{[\phi]}$ are assumed second-order, the analysis of appendix \ref{sec:calc} again gives the effective distributional sources, equations \eqref{eq:Eg1} and \eqref{eq:Ephi1}.  Finally, identical computations based on the linearized distributional Bianchi identity in Fermi normal coordinates give equations \eqref{eq:eom-m-coord} and \eqref{eq:eom-gamma-coord} (as well as $N^{[g]}_{i \mu}=0$).

At this point in the treatment of Einstein-scalar theory the parameters $N^{[g]}_{00}$ and $N^{[\phi]}$ appearing in equations \eqref{eq:eom-m-coord} and \eqref{eq:eom-gamma-coord} were interpreted via an analysis of the specific field equations for Einstein-scalar theory, where it was found that $N^{[g]}_{00} = 4 \pi c_{00} = 8 \pi M$ and $N^{[\phi]}=-16\pi c^{[\phi]} = -16 \pi q$, where $M$ and $q$ are the conventional notions of mass and charge.  For a general theory, the relationship between $\{ N^{[g]}_{00}(t), N^{[\phi]}(t) \}$ and $\{ c_{\mu \nu}(t,\theta,\phi), c^{[\phi]}(t,\theta,\phi) \}$ will depend on the details of the field equations and will in general be more complicated; furthermore, there may not be standard notions of mass and charge available.  I will simply \textit{define} ``mass'' $M=1/8\pi N^{[g]}_{00}$ and ``charge'' $q = -1/16\pi N^{[\phi]}$ for a general scalar-tensor theory.  For any particular theory, one may determine a formula for $M$ and $q$ in terms of  $c_{\mu \nu}$ and $c^{[\phi]}$, enabling the calculation of $M$ and $q$ for a particular body from its exterior field via the appearance of $c_{\mu \nu}$ and $c^{[\phi]}$ in equations \eqref{eq:gbar0} and \eqref{eq:phibar0}.  With $M$ and $q$ incorporated, the results for a general scalar-tensor theory are (\ref{eq:E1ab}-\ref{eq:eom-m}).  

Note that one can easily consider theories with multiple scalar fields as well.  That is, suppose that the Lagrangian of \eqref{eq:scalar-action} depends on a whole collection of scalar fields ${\phi_I}$.  The assumptions are then made for each $\phi_I$, and the steps of the derivation proceed apace, with copies of equations for each $\phi_I$ as well as sums over $I$ where appropriate.  For example, the right-hand-sides of \eqref{eq:es-bianchi}, \eqref{eq:es-bianchi2}, \eqref{eq:E1ab}, \eqref{eq:eom-gamma} and \eqref{eq:eom-m} become sums with one term for each $\phi_I$, while equations (\ref{eq:phi0}-\ref{eq:Ephi1}) and \eqref{eq:E1} are copied for each $\phi_I$.  Thus the results are that there is a charge $q^I$ associated with each scalar field, and that the force on a body is the sum of the ordinary scalar force from each $\phi_I$ (and likewise for the mass evolution).  Note that if one adopts the additional assumption on the field equations for a particular $\phi_I$ that $q^I=0$ for all bodies (i.e., that bodies do not ``produce'' this field), then the field $\phi_I$ has the interpretation of being ``non-interacting'' (except by gravity) and does not appear in the force law.  One can add ``matter fields'' to a theory in this way.

Of course, it is not at all obvious that our assumptions---argued for in the specific context of Einstein-scalar theory (equivalently ``Einstein frame'' scalar-tensor theory)---will remain reasonable for a general theory of the form \eqref{eq:scalar-action}.  I now give some examples of more general theories in which the assumptions do appear to remain reasonable---that is, in which one still expects families of solutions smooth in $\alpha$ and $\beta$ to exist.  The first example is the Einstein-massive-scalar theory, formed by the addition of a mass term to equation \eqref{eq:es-action}, giving the Lagrangian\footnote{Note that the constant $\ell$ has dimensions of length (even if $G \neq 1$); the name ``massive'' for this theory comes from the fact that $\hbar/\ell$ would give the mass of excitations of a quantized $\phi$-field.}
\begin{equation}\label{eq:ems-action}
\mathcal{L} = \sqrt{-g} \left[ R - 2 \left( g^{ab} \nabla_a \phi \nabla_b \phi + \ell^{-2} \phi^2 \right) \right].
\end{equation}
  In this case, rather than the Coulomb potential family $\lambda/r$, the example to keep in mind is the Yukawa potential family $(\lambda/r) e^{-r/\ell}$.  Despite falling off faster than any power of $1/r$ at fixed $\lambda$, the Yukawa potential family is indeed smooth in $\alpha$ and $\beta$; it is simply $\beta e^{-\alpha/\ell}$.  Therefore, the addition of the mass term does not appear to pose any obstacle to smoothness in $\alpha$ and $\beta$.  Notice that the scaled limit $\alpha \rightarrow 0$ gives $\beta = 1/\bar{r}$, reflecting appropriate body-like falloff in the ``buffer zone'', even though this falloff does not occur at fixed $\lambda$.

A second, more complicated example concerns so-called ``chameleon'' theories \cite{khoury-weltman}, in which non-linear effects, as well as non-minimal coupling to matter, cause a body's exterior scalar field---and hence its inferred scalar charge---to depend on the local density of matter.\footnote{A matter field representing ambient density can be included in the action in the manner described in the paragraph above that containing \eqref{eq:ems-action}.}  Although no exact solutions with chameleon behavior are known, approximate solutions that have been compared to numerical solutions \cite{khoury-weltman} show that the exterior field of an isolated body becomes Yukawa if the ambient density is constant (as expected from the linearization of the scalar field equation).  Since $\alpha$ and $\beta$ near $(0,0)$ refers precisely to the ``buffer zone'' where the density of the external universe would be approximately constant, it seems reasonable to expect smoothness to hold here.  In fact, this type of argument should work for any theory with a standard kinetic term, since linearization in the body exterior will give Yukawa.

The equations of motion for screened bodies in chameleon theory were previously investigated by Hui, Nicolis, and Stubbs \cite{hui-nicolis-stubbs}, who obtained the non-relativistic limit of \eqref{eq:eom-gamma} via a variant of the original Einstein-Infeld-Hoffman approach.  We have rigorously derived the full behavior of chameleon bodies, \eqref{eq:eom-gamma} and \eqref{eq:eom-m}, with no non-relativistic approximations.  For a Newtonian body with the scalar field coupling usually \cite{khoury-weltman} considered, the scalar charge $q$ corresponds the body's ``screened mass''.  That is, one has a relationship $q=q(M,\phi^{(0)})$, providing deterministic evolution.  However, it seems unlikely that any such universal relationship will exist for strong-field bodies.  The evolution of the scalar charge $q$ would have to be determined by other methods in order to take advantage of \eqref{eq:eom-gamma} and \eqref{eq:eom-m} for strong-field bodies.

\subsection{Scaling and Universality}\label{subsec:examples}

In this subsection I digress to point out a connection between scaling and universality that arises in this work.  Returning to the example of Einstein-massive-scalar \eqref{eq:ems-action}, recall that the Yukawa potential example family $\beta e^{-\alpha / \ell}$ had the scaled limit $\alpha \rightarrow 0$ of $\beta = 1/\bar{r}$, which is not the Yukawa potential but the Coulomb potential.  The fact that the scaled limit gives a field configuration that is not a solution of the theory can be traced to the theory's lack of \textit{scale invariance} (see appendix \ref{sec:scale}).  In particular, the Lagrangian \eqref{eq:ems-action} does not scale homogeneously under the rescalings $g_{ab} \rightarrow \lambda^2 g_{ab}$ and $\phi \rightarrow \phi$.  Rather, if one rewrites in terms of the barred metric and scalar field, one has
\begin{equation}
\mathcal{L} = \lambda^2 \sqrt{-\bar{g}} \left[ R - 2 \left( \bar{g}^{ab} \nabla_a \bar{\phi} \nabla_b \bar{\phi} + \lambda^2 \ell^{-2} \bar{\phi}^2 \right) \right],
\end{equation}
where $R$ is now constructed from $\bar{g}_{ab}$.  Since the mass term disappears in the $\lambda \rightarrow 0$ limit, the equations satisfied by $\bar{g}_{\bar{\mu} \bar{\nu}}^{(0)}$ and $\bar{\phi}^{(0)}$ are the \textit{massless} Einstein-scalar field equations (a fact easily verified at the level of the equations of motion).  This explains the appearance of the Coulomb potential in the scaled limit of the Yukawa potential example family.

The fact that the body exterior fields $\bar{g}_{\bar{\mu} \bar{\nu}}^{(0)}$ and $\bar{\phi}^{(0)}$ satisfy the massless equations is an indication that our results apply only in situations where the mass term can be neglected in the buffer zone outside a body.  Thus in particular our assumptions require that the body be small compared to $\ell$.  However, this requirement is not surprising or in any sense additional to the basic requirement of the existence of a buffer zone.  Because the scale of variation of solutions to the massive Klein-Gordin equation is rigidly fixed by $\ell$ (in that derivatives\footnote{This language is slightly sloppy.  One really means that scales of variation as measured by the metric, such as $\sqrt{|g^{ab}\nabla_a \phi \nabla_b \phi|}$, are of order $\phi/\ell$.} of $\phi$ are of order $\phi/\ell$), the usual requirement that the body be small compared to the scale of variation of the external fields in fact implies that the body be small compared with $\ell$.  This implication is conveniently captured in the mathematics of the scaled limit.

In a general theory (including the higher-rank theories discussed later) the story will be analogous: by construction, the scaled limit picks out a scale-invariant ``subtheory,'' which must describe the body exterior field approximately for our results to be useful.  Thus one obtains universal behavior only in the presence of scale-invariance, a situation reminiscent of well-known connections between scaling and universality in condensed matter and particle physics.  Note, however, that the field near the body is not required to be described by a scale-invariant theory; our assumptions only refer to $\bar{r} \geq \bar{R}$, and furthermore the results depend only on $\bar{r} \rightarrow \infty$ properties of the body exterior fields.  Scale-invariance appears only in the buffer zone; scale-non-invariant effects (such as ``chameleon effects'' or those due matter fields) are always allowed to act near the body.

\section{Vector-tensor theories}\label{sec:vector}

Vector-tensor theories, the most famous of which is Einstein-Maxwell theory, form another important class of classical field theories.  Here the Lagrangian depends on the metric tensor and a vector field $A^a$,
\begin{equation}\label{eq:vector-action}
S = \int d^4x \mathcal{L}(g^{ab},A_a).
\end{equation}
As in the scalar case we assume that the Lagrangian is diffeomorphism-covariant and the field equation operators $E^{[g]}_{ab} \equiv (-g)^{1/2} \delta S / \delta g^{ab}$ and $E^{[A]a} \equiv (-g)^{1/2} \delta S / \delta A_a$ are second order (local) differential operators.  A useful example to keep in mind is Einstein-Maxwell theory, where (with the normalization of \cite{wald}) one has $E^{[g]}_{ab} = G_{ab}-8\pi T^{EM}_{ab}$ and $E^{[A]b}=-8 \nabla_a \nabla^{[a} A^{b]}$.  Next we derive the Bianchi identity for a vector theory.  Varying \eqref{eq:vector-action} with respect to an infinitessimal diffeomorphism, we have
\begin{equation}
0 = \int d^4x \sqrt{-g} \left\{ E^{[g]}_{ab} (-2\nabla^a \xi^b) + E^{[A]a}\left( \xi^c \nabla_c A_a + A_c \nabla_a \xi^c \right) \right\},
\end{equation}
for a vector field $\xi^a$.  After integration by parts, the fact that $\xi^a$ is arbitrary gives
\begin{equation}\label{eq:vector-bianchi}
\nabla^a E^{[g]}_{ab} = E^{[A]a}\nabla_{[a}A_{b]} + \frac{1}{2} \nabla_a E^{[A]a} A_b .
\end{equation}
As in the scalar case, this identity describes how any ``extra'' stress-energy $E^{[g]}_{ab}$ must be non-conserved in the presence of any ``extra'' charge-current $E^{[A]a}$ for consistent coupling.  Note that this identity may also be derived by varying with respect to the upper-index $A^a$.  (However, the appearance will be different when expressed in terms of $(-g)^{1/2} \delta S / \delta g^{ab}$ computed at fixed $A^a$, since this quantity differs from $E^{[g]}_{ab}$ (computed at fixed $A_a$) by terms proportional to $E^{[A]a}$.)  We restrict without loss of generality to a lowered-index dynamical variable in this section.

An important special case of vector theories are those whose Lagrangian possesses the Maxwell gauge symmetry $A_a \rightarrow A_a + \nabla_a \psi$.  In this case an analogous calculation gives $\nabla_a E^{[A]a}=0$ as an identity (describing the requirement that any ``extra'' charge-current be conserved).  Thus for theories with the gauge symmetry we have two identities,
\begin{align}
\nabla^a E^{[g]}_{ab} & = E^{[A]a}\nabla_{[a}A_{b]} \label{eq:spin1-bianchi} \\ 
\nabla_a E^{[A]a} & = 0. \label{eq:spin1-charge}
\end{align} 

The assumptions for a vector field can be motivated by considering the example of Einstein-Maxwell theory.  Analogously to the scalar case, the sort of behavior we desire is represented by the Coulomb field family $A_0 \sim \lambda/r$.  To characterize this type of behavior with a scaled limit, we must define $\bar{A}_a=\lambda^{-1} A_a$, so that the scaled limit recovers the Coulomb field $\bar{A}_{\bar{0}} \sim 1/\bar{r}$.  This is also the scaling of $A_a$ that leaves the Einstein-Maxwell theory invariant.  The analog of \eqref{eq:compute-scaled} for the components of the vector potential now holds, leading again to our assumptions for each component $A_\mu$ in the coordinates $(t,x^i)$.  Note that our assumptions are on especially strong footing in ordinary electromagnetism, since they were in fact \textit{shown} to hold for the retarded solution of a family of shrinking charge-current and stress-energy sources in flat spacetime \cite{gralla-harte-wald}.  I will make these assumptions for a general theory of the form \eqref{eq:vector-action}.

Therefore the assumptions for this section are the original assumptions of section \ref{sec:GR}, with the metric components replaced by the pair $\{g_{\mu \nu},A_{\mu}\}$, which must satisfy $E^{[g]}_{ab}=0$ and $E^{[A]a}=0$ instead of Einstein's equation.  The computation of the small body equations of motion proceeds in precise analogy with the scalar case of section \ref{sec:scalar}.  That is, define $A^{(0)}_\mu\equiv A_\mu(\lambda=0)$, $A^{(1)}_\mu \equiv \partial_\lambda A_\mu(\lambda)|_{\lambda=0}$, and $\bar{A}^{(0)}_{\bar{\mu}}\equiv \lim_{\lambda \rightarrow 0} \bar{A}_{\bar{\mu}}(\lambda)$ (limit at fixed $\bar{x}^\mu$).  Then the assumptions give series expansions,
\begin{align}
A^{(0)}_\mu & = b^{[A]}_\mu(t) + O(r) \label{eq:A0}\\
A^{(1)}_{\mu} & = c^{[A]}_\mu(t,\theta,\phi)\frac{1}{r} + O(1) \label{eq:A1}\\
\bar{A}^{(0)}_{\bar{\mu}} & = b^{[A]}_\mu(t_0) + c^{[A]}_{\mu}(t_0,\theta,\phi) \frac{1}{\bar{r}} + O \left( \frac{1}{\bar{r}^2}\right), \label{eq:Abar0}
\end{align}
for some $b^{[A]}_\mu$ and $c^{[A]}_\mu$.   The body exterior vector field $\bar{A}^{(0)}_{\bar{\mu}}$ is seen to be stationary and to approach a constant value as $\bar{r}\rightarrow\infty$, confirming its interpretation.  Again one finds the effective distributional sources at linear order,
\begin{align}\label{eq:Eg1A}
E^{[g](1)}_{\mu \nu}[g^{(1)},A^{(1)}] & = N^{[g]}_{\mu \nu}(t) \delta^{(3)}(x^i) \\
E^{[A](1)\mu}[g^{(1)},A^{(1)}] & = N^{[A]\mu}(t) \delta^{(3)}(x^i),\label{eq:EA1}
\end{align}
where $E^{[g](1)}_{\mu \nu}$ and $E^{[A](1)\mu}$ are the linearizations of $E^{[g]}_{\mu \nu}$ and $E^{[A]\mu}$, respectively, off of the background $\{ g^{(0)}_{\mu \nu},A^{(0)}_{\mu}\}$.  At this point it makes sense to treat separately those theories with the Maxwell gauge symmetry and those without.  For those with the symmetry, we have the identities \eqref{eq:spin1-bianchi} and \eqref{eq:spin1-charge}.  Employing as usual the linearized, distributional forms of these identities in Fermi normal coordinates, we find $N^{[g]}_{i0}=N^{[g]}_{ij}=N^{[A]}_i=0$ as well as $\partial_0 N^{[g]}_{00} = \partial_0 N^{[A]}_0 = 0$ and $M a_i = N^{[A]0} \partial_{[0}A^{(0)}_{i]}$.  As usual $N^{[g]}_{00}$ and $N^{[A]}_0$ may be interpreted by their appearance in the body exterior metric and vector field.  In the Einstein-Maxwell case discussed above one sees that the usual notions of mass $M$ and charge $q$ are related by $N^{[A]0}=16 \pi q$ and $N^{[g]}_{00}=8 \pi M$.  We use this to define $q$ and $M$ for a general theory with Maxwell gauge invariance, and rewrite the results covariantly to obtain
\begin{align}
E^{[g](1)}_{ab} & = 8 \pi M \int_\gamma u_a u_b \delta_4(x,z(\tau)) d\tau \label{eq:E1abA}\\
E^{[A](1)a} & = 16 \pi q \int_\gamma u^a \delta_4(x,z(\tau)) d\tau , \label{eq:E1A}
\end{align}
as well as
\begin{align}
M u^b \nabla_b u_a & = q u^b (2 \nabla_{[a}A^{(0)}_{b]}), \label{eq:lorentz}
\end{align}
where $q$ and $M$ are constants.  Thus for theories with the Maxwell gauge symmetry we have the usual point particle stress-energy and charge-current, along with the Lorentz force law.  For theories without the gauge symmetry, however, the situation is more complicated.  In this case we only have the single identity  \eqref{eq:vector-bianchi}, and the Fermi coordinate calculation now gives
\begin{align}
N^{[g]}_{i\mu} & = \frac{1}{2} N^{[A]}_{\ \ \ i} A^{(0)}_\mu \label{eq:c1} \\
\partial_0 N^{[g]}_{00} & = -N^{[A]\mu} \partial_{[\mu}A^{(0)}_{0]} + \frac{1}{2}A^{(0)}_0 \partial_0 N^{[A]}_{\ \ \ 0} \label{eq:c2} \\
N^{[g]}_{00} a_i & = N^{[A]\mu} \partial_{[\mu}A^{(0)}_{i]} + \frac{1}{2} \partial_0(N^{[A]}_i A^{(0)}_0)-\frac{1}{2}A^{(0)}_i \partial_0 N^{[A]}_0. \label{eq:c3}
\end{align}
Since we no longer have $N^{[g]}_{i0}=N^{[g]}_{ij}=N^{[A]}_i=0$, the parameters  $q$ and $M$ no longer suffice to characterize the body.  In light of equation \eqref{eq:c1}, which requires $N^{[A]}_{ \ \ \ i}$ to point along $A^{(0)}_i$, it seems simplest to introduce a second charge $\hat{q}$ by $N^{[A]}_{\ \ \ i}=16 \pi \hat{q} A^{(0)}_i$.  Then the results are rewritten covariantly as
\begin{align}
E^{[g](1)}_{ab} & = 8 \pi \int_\gamma \left( M u_a u_b + 4 \hat{q} P^c_{\ (a}A_{b)} A^{(0)}_c \right) \delta_4(x,z(\tau)) d\tau \label{eq:p1}\\
E^{[A](1)a} & = 16 \pi \int_\gamma \left( q u^a + \hat{q} P^{ab} A^{(0)}_b\right) \delta_4(x,z(\tau)) d\tau \label{eq:p2} 
\end{align}
 and
\begin{align}
\left( M - \hat{q}(A^c A_c)^2 \right) u^a \nabla_a u_b & = 2 (q - \hat{q} A^c u_c ) \nabla_{[b}A_{a]} u^a \nonumber \\
& \quad + P^a_{\ \ b} \left\{ 2 \hat{q} A^c\nabla_{[c} A_{a]} + u^c \nabla_c \left( \hat{q} A_d u^d A_a \right) - A_a u^c \nabla_c q \right\} \label{eq:p3} \\
u^a \nabla_a M & = -\hat{q} A^a u^b(2 \nabla_{[a}A_{b]}) + u^a A_a u^b\nabla_b q, \label{eq:p4}
\end{align}
where $P^a_{\ b} \equiv \delta^a_{\ b} - u^a u_b$ projects orthogonally to $u^a$, and the superscript $(0)$ on $A^{(0)}_a$ has been dropped in the last two equations for readability.
Thus for theories without the gauge symmetry the usual point particle stress-energy and charge-current are \textit{not} obtained (in that the distributional forms are not parallel to $u^a$), the Lorentz force law is \textit{not} the correct force law, and the particle is described by a time-dependent mass $M$ as well and two charges $q$ and $\hat{q}$, neither of which has an evolution law (just as there was no law for the scalar charge in scalar-tensor theory).  An example of a commonly studied theory to which these equations apply is Einstein-Proca theory.  Of course, in most references Proca lagrangian is coupled to matter via an interaction term that by itself has the gauge symmetry, so that solutions to the Proca-matter system respect charge conservation (but charge conservation does \textit{not} hold as an identity).  If one restricts to such matter, presumably one would have $\hat{q}=0$ and $u^a \nabla_a q = 0$ for bodies made of that matter, whence the usual particle equations of electromagnetism (\ref{eq:E1abA}-\ref{eq:lorentz}) would be recovered.  However, it is not clear that there is reason for matter to conserve charge in a vector theory beyond the fact that it is required in the most familiar vector theory.  It should be emphasized that the Proca theory (and others without the gauge symmetry) admit far more general behavior in the motion of bodies than does ordinary electromagnetism.  This general behavior would give the motion of any matter that did not conserve charge, as well as the motion of any non-matter objects (such as black holes or ``geons'') that might exist in the theory.

Note finally that the analysis of this section can be straightforwardly generalized to the case of multiple vector fields (or even multiple vector and scalar fields) in the manner discussed in section \ref{sec:scalar} for scalar-tensor theory.  In this case one simply obtains copies of equations \eqref{eq:p1} and \eqref{eq:p2} for each field, and the right-hand-sides of \eqref{eq:p3} and \eqref{eq:p4} are copied for each field to form a sum.  If any of the fields have Maxwell gauge-invariance, of course, the simpler terms from equations (\ref{eq:E1abA}-\ref{eq:lorentz}) may be used for that field.  There may also be different gauge symmetries that provide different simplification.  An important example is non-Abelian gauge theory, where the charges respect ``gauge covariant'' conservation.  More precisely, if we label the set of vector fields by $A_a^I$ and their field equation operators by $E^{Ia} \equiv (-g)^{-1/2} \delta S / \delta A_a^I$ (using capital Latin indices for ``gauge indices'') then the gauge symmetry of the Lagrangian gives $\nabla_a E^{Ka} = \sum_{I,J} f^{IJK} E^{Ia} A^J_a $ as an identity, where $f^{IJK}$ are the structure constants as defined in \cite{peskin-schroeder} (anti-symmetric on the first two indices).  The usual Fermi normal coordinate calculation on this identity implies that the ``hatted charge'' vanishes for each body and gives an evolution law for the charges $q^I$.  The small body equations of motion are then
\begin{align}
M u^b \nabla_b u_a & = \sum_I q^I (2 \nabla_{[a}A^I_{b]}) u^b \\
u^a \nabla_a M & = \sum_{I,J,K} f^{IJK} q^I A^J_a A^K_b u^a u^b \\
u^a \nabla_a q^K & = \sum_{I,J} f^{IJK} q^I A^J_a u^a .
\end{align}
The evolution is now fully deterministic on account of the extra symmetry of the Lagrangian.  Note that if the structure constants are totally anti-symmetric, then $M$ is constant and these reduce to ``Wong's equations'' \cite{wong}.

\section{The General Case}\label{sec:general}

The procedure used for scalar and vector fields generalizes straightforwardly to higher-rank tensor fields.  Here we motivate the scaling by analogy with the scalings used before.  The scalings used for the metric, scalar, and vector fields were all such that the power of $\lambda$ cancelled powers of $\lambda$ resulting from the Jacobian of the coordinate transformation to scaled coordinates, allowing \eqref{eq:compute-scaled} and its scalar and vector analogs to hold.  For a general-rank tensor field $T^{a_1 \dots a_n}_{\ \ \ \ \ \ \ b_1 \dots b_m}$, I define the scaled version $\bar{T}^{a_1 \dots a_n}_{\ \ \ \ \  \ \ b_1 \dots b_m} = \lambda^{n-m} T^{a_1 \dots a_n}_{\ \ \ \ \ \ \ b_1 \dots b_m}$  so that the analog of \eqref{eq:compute-scaled} holds.  Then the usual reasoning leads to the usual assumptions for each component of the tensor field, and one proceeds exactly as in the previous sections.  I will summarize this procedure in the form of a proof of a theorem, below.  It is straightforward to follow the steps to determine the force law for any theory of particular interest (although higher-rank fields are more seldom considered).

\begin{thm}
Let $S$ be an action in four spacetime dimensions in the sense of appendix E of Wald \cite{wald} such that 1) the Lagrangian $\mathcal{L}$ depends differomorphsim-covariantly on the metric $g_{ab}$ and some set of tensor fields $\{\psi_I\}$ (tensor indices suppressed), i.e., $\phi^*\mathcal{L}(g_{ab},\psi_I) = \mathcal{L}(\phi^*g_{ab},\phi^*\psi_I)$ is satisfied for diffeomorphisms $\phi$; and 2) $E^{[g]}_{ab} = (-g)^{-1/2} \delta S / \delta g^{ab}$ and $E^I = (-g)^{-1/2} \delta S / \delta \psi_I$ are (local) second-order differential operators on $\{g_{ab},\psi_I\}$.  Suppose there exists a one-parameter-family $\{g_{ab}(\lambda),\psi_I(\lambda)\}$ satisfying the analogs of the assumptions of section \ref{sec:GR}.\footnote{That is, adjoin the components of the $\psi_I$ and $\bar{\psi}_I$ to those of the metric and rescaled metric (respectively) where they appear, and replace satisfaction of the Einstein equations with satisfaction of the field equations $E^{[g]}_{ab}=0$ and $E^I=0$.}  Then, the worldline $\gamma$ (four-velocity $u^a$ and four-acceleration $a^a$) and a certain function $M$ defined on $\gamma$ satisfy equations of the form $M a^a=f^a$ and $u^a \nabla_a M = F$, where $f^a u_a=0$ and both $f^a$ and $F$ are local tensor functions of $u^a$, $a^a$, $g_{ab}(0)|_\gamma$, $\psi_I(0)|_\gamma$, $\nabla (\psi_I(0))|_\gamma$, and certain tensor fields $q^I$ defined on $\gamma$.
\end{thm}

The proof is essentially to follow the steps of the previous sections.  Since these steps are by now familiar, I will omit some details in the description here (allowing considerable savings on notation).  Begin by varying $S$ with respect to an infinitessimal diffeomorphism to derive the ``Bianchi identity''.  This results in an expression of the form
\begin{equation}\label{eq:general-bianchi}
\nabla^a E^{[g]}_{ab} = \sum_I \left[ \left( E^I \odot \nabla \psi_I \right)_b + \left( \nabla E^I \odot \psi_I \right)_b \right]
\end{equation}
where the notation $\left( A \odot B \right)_a$ indicates a sum of terms, each of which consists of the tensor product of $A$ and $B$ contracted in some way to yield a dual vector.  Note that the explicit form for an arbitrary-rank tensor field was worked out in \cite{seifert-wald}.  Now derive the effective point particle description.  By assumption iii), the components of $g^{(1)}_{ab}$ and $\psi^{(1)}_I$ are $O(1/r)$ ($r$ near zero).  By the assumption of second-order field equations and the analysis of appendix \ref{sec:calc}, the components of $E^{[g](1)}_{ab}$ and $E^{I(1)}$ as distributions are multiples of $\delta^3(x^i)$; take the coefficients to be $N^{[g]}_{\mu \nu}(t)$ and $N^I(t)$ (component indices suppressed), respectively.  Now apply the linearized, distributional form of equation \eqref{eq:general-bianchi}.  By satisfaction of the background field equations ($E^{(0)}_{ab}=E^{I(0)}=0$), this identity takes the form
\begin{equation}
\nabla^\mu E^{[g](1)}_{\mu \nu} = \sum_I \left[ \left( E^{I(1)} \odot \nabla \psi^{(0)}_I \right)_\nu + \left( \nabla E^{I(1)} \odot \psi^{(0)}_I \right)_\nu \right] ,
\end{equation}
where here and below $\nabla$ is the derivative operator associated with the background metric.  We then have
\begin{equation}
\nabla^\mu N^{[g]}_{\mu \nu} = \sum_I \left[ \left( N^{I} \odot \nabla \psi^{(0)}_I|_\gamma \right)_\nu + \left( \nabla N^{I} \odot \psi^{(0)}_I|_\gamma \right)_\nu \right].
\end{equation}
Adopting Fermi normal coordinates, one has
\begin{align}
\delta^{(3)}(\vec{x})\left[ - \partial_0 N_{00} + a_i N_{i0} - \mathcal{F} \right] + \partial_i \delta^{(3)}(\vec{x}) \left[ N_{i0} - \mathcal{G}_i \right] & = 0 \label{eq:FG} \\
\delta^{(3)}(\vec{x})\left[ - \partial_0 N_{0i} + a_j N_{ij} + a_i N_{00} - \mathcal{H}_i \right] + \partial_i \delta^{(3)}(\vec{x}) \left[ N_{ij} - \mathcal{K}_{ij} \right] & = 0, \label{eq:HK}
\end{align}
where $\mathcal{F}$, $\mathcal{G}_i$, $\mathcal{H}_i$, and $\mathcal{K}_{ij}$ are local functions of $N^I$, $\psi^{(0)}_I|_\gamma$, $\partial \psi^{(0)}_I|_\gamma$, $a_i$.  Naming $N_{00}=M$, it now follows from \eqref{eq:FG} and \eqref{eq:HK} that
\begin{align}
\partial_0 M & = a_i \mathcal{G}_i + \mathcal{F} \label{eq:mass} \\
M a_i & = \partial_0 \mathcal{G}_i - a_j \mathcal{K}_{ij} + \mathcal{H}_i .\label{eq:accel} 
\end{align}
These are the small-body equations of motion for the theory (expressed in Fermi normal coordinates).  It is straightforward to determine the functions $\mathcal{F}$, $\mathcal{G}_i$, $\mathcal{H}_i$, and $\mathcal{K}_{ij}$ by direct calculation for the theory in question (or even the general case; however, the expression is not simple).  It is also straightforward to relate the parameters $M$ and $N^I$ to the body exterior configuration $\{\bar{g}^{(0)}_{\bar{\mu} \bar{\nu}},\bar{\psi}^{(0)}_i\}$ for each particular theory (as done in previous sections), enabling their calculation for any particular body.  The covariant translation of equations \eqref{eq:mass} and \eqref{eq:accel} proves the theorem.

The theorem establishes that a simplified description of motion (in the form of a second-order equation for $\gamma$)\footnote{In some theories the field equations may enforce $M=0$ in which case the equation for $\gamma$ may be lower than second-order or even trivial.} is obtained via buffer-zone dynamics in a very large class of theories. It is interesting to speculate on the extent to which the class of theories could be enlarged.  The requirement of second-order field equations seems easiest to relax, since higher-order equations would lead only to higher-order delta functions (i.e., derivatives of delta functions) appearing in the effective stress-energy, whence our calculations could proceed straightforwardly.  However, it is far from obvious that our assumptions on one-parameter-families would remain reasonable in the context of higher-order theories, whose solutions may have very different properties.\footnote{Note, however, that many higher-order theories (such as those whose Lagrangian is a function of the Ricci scalar) admit second-order formulations, so that the current analysis applies.} The requirement of a diffeomorphism-covariant Lagrangian appears difficult if not impossible to relax, since the Bianchi identity plays an essential role in determining the motion.  At least in our approach, the diffeomorphism-symmetrry and Lagrangian formulation are key to obtaining a description of motion from the buffer-zone field equations alone.  

\section{Summary}

I have treated the motion of small bodies in classical field theory via the approach of \cite{gralla-wald}.  The search for one-parameter-families of solutions representing the exterior field of a shrinking body led precisely to the physical assumption of a ``buffer zone''---a region far enough from the body that its field can be approximated in a multipole series, but close enough to the body that the field of the external universe can be approximated in an ordinary Taylor series.  No assumptions about the body interior are made.  In the case of second-order metric-based theories following from a diffeomorphism-covariant Lagrangian, I derived the force law for scalar and vector fields, and showed that the method works in the general-rank case.  This provides a rigorous derivation of the small-body force law in many classical field theories commonly considered, and shows that field dynamics outside a body determines its motion in a very general class of theories.

\section*{Acknowledgements}
I wish to thank Wayne Hu, Fabian Schmidt, Michael Seifert and especially Bob Wald for very helpful conversations.  This research was supported in part by NSF grants PHY04-56619 and PHY08-54807 to the University of Chicago.

\appendix

\section{Delta-function Calculation}\label{sec:calc}

Consider a linear, second-order differential operator $L$ that takes tensors $\phi^{a_1...a_n}$ of rank $n$ into tensors $(L\phi)^{b_1...b_m}$ of rank $m$.  The adjoint $L^\dagger$ is the linear map from tensors of rank $m$ to tensors of rank $n$ defined by \cite{wald-prl} 
\begin{equation}\label{eq:adjoint}
\psi^{b_1...b_m} (L \phi)_{b_1...b_m} - (L^\dagger \psi)^{a_1...a_n} \phi_{a_1...a_n} = \nabla_c s^c
\end{equation}
for arbitrary $\psi$ and $\phi$.  The vector $s^c$ is a multilinear function of $\phi^{a_1...a_n} \nabla_c \psi^{b_1...b_n}$,  $\psi^{b_1...b_n} \nabla_c \phi^{a_1...a_n}$, and $\phi^{a_1...a_n}\psi^{b_1...b_n}$.  We now promote $L$ to an operator on distributions in the standard way.  That is, we define a distributional operator $\hat{L}$ by
\begin{equation}\label{eq:Lhat}
(\hat{L}\phi)[f] \equiv \int d^4x  \sqrt{-g} (L^\dagger f)^{a_1...a_n}\phi_{a_1...a_n}
\end{equation}
for smooth test tensors $f^{b_1...b_m}$ of compact support.  Now suppose that the metric and $\phi$-field have the following expressions in coordinates $(t,x^i)$,
\begin{align}
g_{\mu \nu} & = \eta_{\mu \nu} + O(r) \label{eq:gform} \\
\phi^{\mu_1...\mu_n} & = \frac{1}{r}C^{\mu_1...\mu_n}(t,\theta,\phi) + O(r^0), \label{eq:phiform}
\end{align}
and further that $\phi^{\mu_1...\mu_n}$ solves the equation at $r>0$, i.e., that we have
\begin{equation}\label{eq:soln}
(L \phi)^{\nu_1...\nu_m} = 0, \ \ \ r>0 .
\end{equation}
Here $r,\theta,\phi$ are related to $x^i$ in the usual way.  The distribution $(\hat{L} \phi)^{\nu_1...\nu_m}$ may be computed by
\begin{align}
(\hat{L}\phi)[f] & = \lim_{\epsilon \rightarrow 0} \int_{\epsilon>0} d^4x  \sqrt{-g} (L^\dagger f)^{\mu_1...\mu_n}\phi_{\mu_1...\mu_n} \\
& = \lim_{\epsilon \rightarrow 0} \int_{\epsilon>0} d^4x  \sqrt{-g} \left\{ f^{\nu_m...\nu_m} (L \phi)_{\nu_1...\nu_m} + \nabla_\rho s^\rho \right\} \\
& = \lim_{\epsilon \rightarrow 0} \int_{r=\epsilon} r^2 \sin \theta d\theta d\phi dt \ n_\rho s^\rho,
\end{align}
where $n_\rho$ is the unit normal to the unit two-sphere in Euclidean three-space, and $s^\rho$ is evaluated on $\phi$ and $f$.  The first line follows from \eqref{eq:Lhat} and the smoothess of $f^{\mu_1...\mu_n}$.  The second line follows from \eqref{eq:adjoint}.  The third line follows from \eqref{eq:soln} and integration by parts, where the boundary term vanishes by the compact support of $f^{\mu_1...\mu_n}$.  The volume element on the surface $r=\epsilon$ has been replaced with the Minkowski volume element on account of the limit $\epsilon \rightarrow 0$ and the metric form \eqref{eq:gform}.  (Note that if different coordinates were chosen such that the metric is not Minkowski at $r=0$, the volume element would take a different form, but it is easy to check that the analysis would still hold.)  Now, equation \eqref{eq:phiform} together with the properties of $s^a$ imply that $s^\rho$ may be written 
\begin{equation}
s^\rho=\frac{1}{r^2} D^{\rho \nu_1...\nu_m}(t,\theta,\phi)f_{\nu_1...\nu_m}(t,r,\theta,\phi) + O(1/r)
\end{equation}
for some $D^{\rho \nu_1...\nu_m}$.  Then we have
\begin{align}
(\hat{L}\phi)[f] & = \lim_{\epsilon \rightarrow 0} \int \sin \theta d\theta d\phi dt \ n_\rho  D^{\rho \nu_1...\nu_m}f_{\nu_1...\nu_m}|_{r=\epsilon} \\
& = \int dt f_{\nu_1...\nu_m}|_{r=0} \int \sin \theta d\theta d\phi \ n_\rho  D^{\rho \nu_1...\nu_m} \\
& \equiv \int dt N^{\nu_1...\nu_m}(t) f_{\nu_1...\nu_m}(t,x^i=0) 
\end{align}
where the second step follows from the smoothness of $f^{\nu_1...\nu_m}$, and the last step simply defines the result of the angular integral to be $N^{\nu_1...\nu_m}(t)$.  Since the test function is evaluated at $x^i=0$, this expression shows that distribution $(\hat{L} \phi)^{\nu_1...\nu_m}$ is proportional to the spatial delta function $\delta^{(3)}(\vec{x})$; i.e., we have the desired result
\begin{equation}
(\hat{L} \phi)^{\nu_1...\nu_m} = N^{\nu_1...\nu_m}(t) \delta^{(3)}(\vec{x}).
\end{equation}
Note that an explicit formula for $N^{\nu_1...\nu_m}(t)$ in terms of $C_{\mu_1...\mu_n}(t,\theta,\phi)$ can be determined for any particular differential operator $L$ by following the steps of this computation explicitly.  This formula will in general involve angular integrals of $C_{\mu_1...\mu_n}$ and its first angular derivatives.

\section{Scale Invariance}\label{sec:scale}

In this appendix I provide a definition of scale invariance and compare with other definitions commonly given.  Tensor indices are suppressed throughout this section.  It is convenient to restrict to theories that follow from a Lagrangian,
\begin{equation}
S(g,\psi_I) = \int d^4x \mathcal{L}(g,\psi_I),
\end{equation}
although the metric need not be dynamical and there may exist other ``background structure''.  (That is, the Lagrangian is not required to depend diffeomorphism-covariantly on $g$ and $\psi_I$.)
The theory is scale invariant if the action scales homogeneously under a scaling of the metric and other fields; more precisely, if exist numbers $\{P_I\}$ and $n$ such that
\begin{equation}
S(\lambda^2 g, \lambda^{P_I} \psi_I) = \lambda^n S(g,\psi_I),
\end{equation}
for numbers $\lambda$.  This property implies that the equations of motion for the rescaled fields are identical to the equations of motion for the original fields, so that there is no preferred scale for lengths (as measured by $g$) or field values ($\phi$) in the theory.  The choice of $\lambda^2$ for the scaling of $g$ is conventional, and gives $\lambda$ the interpretation of a length (since the line element scales as $\lambda^2$).  Note that our definition of $\bar{g} = \lambda^{-2} g$ in the body of the paper is consistent with this convention of ``$g \rightarrow \lambda^2 g$''.

An alternative definition of scale-invariance is often given in the context of theories specified by partial differential equations in coordinates (without the introduction of a metric).  In this case one rescales the coordinates $x^\mu \rightarrow \lambda x^\mu$ and asks if a rescaling of the fields can restore the original partial differential equations.  Such theories can often be rewritten diffeomorphism-covariantly via the introduction of a flat metric $g_{ab}$.  Changing the coordinates of the original partial differential equation then corresponds to applying the diffeomorphism $\sigma$ associated with coordinate rescaling to all fields \textit{except} the metric.  That is, to check if a diffeomorphism-covariant equation $E[g,\psi_I]=0$ is scale-invariant according to the non-tensorial definition, one asks if $\sigma_*E[g,\psi_I] = \lambda^n E[g,\lambda^{P_I} \sigma_*\psi_I]$.  Applying $\sigma^{-1}$ to both sides and using the diffeomorphism-covariance of $E$, this becomes $E[g,\psi_I] = \lambda^n E[\sigma^{-1}_{\ \ *}g,\lambda^{P_I}\psi_I]$.  But since $g$ is flat, $\sigma^{-1}_{\ \ *}g = \lambda^2 g$, and this reduces to the definition involving rescaling the metric (at the level of the field equations).  This is why one says that rescaling the metric is a curved-spacetime generalization of rescaling the coordinates \cite{hollands-wald}.

In the context of quantum field theory a second alternative definition of classical scale-invariance is often given by requiring that the action be left \textit{invariant} under the rescalings, rather than just scale homogeneously \cite{CFT}.  This corresponds to our definition with the additional demand that $n=0$.  Since the overall scaling of the action (i.e., the value of $n$) has no effect on the classical equations of motion, this notion in fact removes as ``classically scale-invariant'' many classical theories with no preferred scale.  For example, vacuum general relativity is scale-invariant in our sense, having no preferred length scale; however, vacuum general relativity is not classically scale invariant according to the definition used in quantum field theory.  Another example is massless $\phi^n$ ($n\neq 2$) in flat spacetime (scale-invariant in our sense); only $\phi^4$ is classically scale-invariant in the sense used in quantum field theory.

The difference between the definitions can be further elucidated with reference to the well-known fact that a scale-invariant action contains only coupling constants that are dimensionless in particle physics units ($c=1$, $\hbar=1$, $G \neq 1$).  Adopting the viewpoint that masses and lengths are fundamentally different (but that time intervals and lengths are not), it is convenient to work in special relativity units ($c=1$,$\hbar \neq 1$,$G \neq 1$), where the statement of dimensionlessness in particle physics units becomes the property of having equal mass and length dimension, so that in particular one cannot construct a length using $\hbar$ (which has dimensions of mass times length).  For example, the constant $\Lambda$ of $\Lambda \phi^4$ has dimensions of mass times length so that no length can be constructed with $\hbar$.  On the other hand, the constant $G$ of general relativity has dimensions of mass over length, so that $\hbar$ can be used to construct a length (called the Planck length).  While our notion of classical scale-invariance implies that no lengths can be constructed from the coupling constants alone, the quantum field theory notion of classical scale-invariance places the further restriction that no lengths can be constructed even when one is allowed to use $\hbar$ in addition.

\end{document}